\documentstyle[aps,epsf,floats]{revtex}
\begin{document}
\author{
 Hans-J\"{u}rgen Sommers$\S$, Yan V. Fyodorov$\S,\P$,
and Mikhail Titov$\P$}

\title{S-matrix poles for chaotic quantum systems as eigenvalues
of complex symmetric random matrices: from isolated to overlapping resonances}

\address{
$\S$Fachbereich Physik, Universit\"at-GH Essen,
D-45117 Essen, Germany
        }
\address{
$\P$  Petersburg Nuclear Physics Institute, Gatchina
188350, Russia }

\date{\today}

\maketitle
\begin{abstract}
We study complex eigenvalues of
large $N\times N$ symmetric random matrices of
the form ${\cal H}=\hat{H}-i\hat{\Gamma}$, where both $\hat{H}$
and $\hat{\Gamma}$ are
real symmetric, $\hat{H}$ is random Gaussian and $\hat{\Gamma}$
is such that
$N\mbox{Tr}  \hat{\Gamma}^2_2\sim \mbox{Tr} \hat{H}_1^2$ when
$N\to \infty$.
When  $\hat{\Gamma}\ge 0$ the model can be used to describe
the universal statistics of S-matrix poles (resonances) in the complex
energy plane. We derive the ensuing distribution of the
resonance widths which generalizes the well-known $\chi^2$ distribution
to the case of overlapping resonances.
 We also consider
 a different class of "almost real" matrices when
$\hat{\Gamma}$ is random and uncorrelated with $\hat{H}$.

\end{abstract}

As is well-known, S-matrix poles or {\it resonances} can be looked at
as complex eigenvalues of an effective
non-Hermitian Hamiltonian ${\cal H}$ emerging after the so-called
{\it complex rotation} was applied to the original Hermitian Hamiltonian
of the system. The method\cite{CC} allows one to extract resonance positions $E_k$
and widths $\Gamma_k$ directly, without expensive evaluation of the
energy-dependent $S-$ matrix elements. The search of resonances
typically amounts to diagonalizing large complex symmetric matrices
representing the effective Hamilonian in a suitable basis on a finite
grid. Using this method, patterns consisting of hundreds
of individual resonances
were extracted recently for realistic models
in atomic and molecular physics\cite{Main,Blumel}. Frequently,
for high enough excitation energies the S-matrix poles are placed
irregularly in the complex plane, forming
 a structure of a "chaotic jumble" \cite{Blumel},
the situation calling for a statistical description. As another important
development it is appropriate to mention a powerful
numerical algorithm that was proposed to extract resonance
positions and widths with a high accuracy from the experimentally
accessible time correlation function of finite duration\cite{Man}.
All these facts make the problem of statistical description of
resonances to be an important and opportune task.

At the same time, the statistics of highly excited
{\it bound} states of  closed quantum systems
of quite different microscopic nature is known to be system-independent
(universal), provided
the corresponding classical counterparts demonstrate a well-developed
chaotic motion  \cite{Bohigas}.
Moreover, the
spectral correlations turn out to
be exactly the same as provided by the theory of
large random matrices on the scale determined by a typical
separation $\Delta$ between neighbouring eigenvalues.
 Microscopic justifications of the use of random matrices
for describing the universal properties of quantum chaotic systems have been
provided by several groups recently, based both on traditional semiclassical
periodic orbit expansions \cite{per,bogomol} and on advanced
field-theoretical methods \cite{MK,aaas}. It is natural to try to
develop a description of universal
statistical properties of resonances using the same ideas.

The methods to adjust random matrix description
to the case of open chaotic systems are well
known since the pioneering paper \cite{VWZ}
and described in much detail in \cite{FSR}.
The starting point of this approach is the representation of the
scattering matrix $\hat{S}$ in terms
of an effective non-Hermitian Hamiltonian
${\cal H}_{eff}=\hat{H}-i\hat{\Gamma}$.
Here the Hermitian Hamiltonian  $\hat{H}$ describes the {\it closed}
counterpart of the open system whereas the
anti-Hermitian part $i\hat{\Gamma}$
arises due to coupling to open scattering channels.
It  has to be
chosen in the form  ensuring the unitarity
of the scattering matrix. As a result
all eigenvalues of $\hat{\Gamma}$ turn out to be non-negative
(as is also required by causality),
the number of strictly positive eigenvalues being just the number
$M$ of open channels.

It is natural to expect, that statistical properties of
resonances are inherited from the corresponding
random matrix universality of levels of closed systems.
Remembering the discussion above,  we expect them
to be universal on the energy scales in the complex plane
comparable with the mean level spacing for the closed system $\Delta$.
In contrast,
properties on a much larger scale must be highly system-specific.

A natural way to incorporate the random matrix
description of the quantum chaotic system is to replace
$\hat{H}$ by a large $N\times N$
random matrix of appropriate symmetry. Namely,
chaotic systems with preserved time-reversal invariance (TRI) should be
described by matrices $H_{ij}$ which are real symmetric. Such
matrices form the Gaussian Orthogonal Ensemble (GOE), whereas for
systems with broken TRI one uses complex Hermitian matrices from
the Gaussian Unitary Ensemble (GUE)\cite{Bohigas}. As to the matrix
$\hat{\Gamma}$ it is determined by the coupling of the chaotic system to
open channels and can be considered as a fixed given one of the
same dimension.

In general, the two matrices $\hat{H}$ and $\hat{\Gamma}$ do not
commmute, the fact making the analysis of complex eigenvalues of
${\cal H}$ to be a rather non-trivial problem. In the simplest case
one can treat the anti-Hermitian part as a small perturbation
to the Hermitian one. This is justified for the regime
of isolated resonances, i.e.
when typical resonance widths are much
smaller than the mean separation
$\Delta$ between the positions of neighbouring resonances.
Assuming for simplicity that all scattering channels (hence, the
eigenvalues of $\hat{\Gamma}$) are statistically equivalent, one
arrives at the so-called $\chi^2$ distribution of the resonance
widths. It was introduced originally many years ago by Porter and
Thomas and since then experimentally observed in many physical
systems, see some references in \cite{Gasp}. The corresponding perturbative
expression is known also for the case of non-equivalent
scattering channels \cite{AL}.

Until quite recently relatively little was known about resonance statistics
non-perturbatively. Among the known facts were the joint probability
density of all resonances (however, not its moments)
 for the systems with only one open channel $M=1$ \cite{Sok}, and the mean
density of complex eigenvalues for very many open channels $M\sim
N\gg 1$\cite{Haake}. The most essential progress was achieved
recently \cite{FS,FSR}
by employing a powerful field-theoretical method known as Efetov's
supersymmetric non-linear $\sigma-$model\cite{Efbook}. It turns out
to be possible to find explicitly the mean density of the resonances
for systems with broken time-reversal invariance and to put
forward a well-grounded conjecture on the higher correlation
functions in the complex plane\cite{FTS}. Unfortunately, the most
interesting and practically important case of the systems with
preserved time-reversal invariance proved to be less accessible
 analytically (for numerics see \cite{reso})
and no results of such generality were obtained for such systems so far.

The main goal of the present paper is to fill in this gap partly
by presenting the exact non-pertubative distribution of the resonance
widths for quantum chaotic systems with preserved time-reversal
invariance.

It is necessary to menion that
 random matrices (more generally, random linear operators)
with complex eigenvalues emerged recently in many other physical contexts.
 A few recent examples include the description of thermal motion of an isolated
vortex in disordered type-II superconductors with columnar defects
, the problem of a
 classical diffusing particle advected
by a stationary random velocity field and  attempts to understand
the universal features of chiral symmetry breaking in Quantum
Chromodynamics (a more detailed discussion
and the relevant references can be found in the papers mentioned below).
 At the same time, our knowledge of statistical
properties of random non-selfadjoint matrices is
quite scarce and incomplete. This fact recently
stimulated efforts of different groups to improve our understanding in this
direction\cite{QCD,Khor,FKS1,Kus,FKS3}.
In particular, the existence of a nontrivial regime
of {\it weak non-Hermiticity} was recognized in the work \cite{FKS1},
see more detailed discussion in \cite{FKS3}.
The guiding idea to realize the existence of such a regime
comes from the experience with resonances\cite{FS}. Guided by that example
one guesses that a new regime occurs when the
imaginary part of typical eigenvalues is comparable
with the mean {\it separation} between
neighboring eigenvalues along the real axis.

One can use again the supersymmetry approach  to obtain the mean density
of complex eigenvalues in the regime of weak non-Hermiticity
for matrices with independent elements \cite{FKS1,FKS3}.
The density turned out to be described by a
formula containing as two opposite limiting cases both the Wigner
semicircular density of real eigenvalues of the Hermitian matrices
and the uniform density of complex eigenvalues discovered
for usual non-Hermitian random matrices already by Ginibre \cite{Gin},
in much details addressed by Girko \cite{Gir} and other authors
\cite{neural}.
 In his insightful paper Efetov \cite{Efnonh} managed to derive
the density of complex eigenvalues for a related, but
different set of (almost-symmetric) real random
matrices. His result was generalized in \cite{FTS}.
Actually, the non-Hermitian matrices considered in \cite{FKS1}
and \cite{Efnonh} are just two limiting cases of a general three-parameter
family of "weakly non-Hermitian" ensembles\cite{FKS3} with
independent entries.
Apart from these two cases,
the mentioned three-parameter family contains one more important particular
case corresponding to an ensemble of complex symmetric random
matrices. Such matrices are natural to call the
"almost-real" ones. The second goal of the present paper is to
provide the eigenvalue density for this particular ensemble.

As a starting point of our investigation we use the
 representations of the mean density $\rho(X,Y)$ of eigenvalues $Z=X+iY$
with a given imaginary part $Y$ derived by us earlier\cite{FKS3,my}.
Let us first consider the case relevant to the chaotic scattering.
Due to rotational invariance of the GOE matrices the result can be dependent
only on the set of positive eigenvalues $\gamma_1,\gamma_2,...,\gamma_M$ of
the matrix $\hat{\Gamma}$.
Correspondingly,
the density $\rho_X(y)
\equiv\rho(X,Y)\Delta^2(X)/\pi$ of scaled
resonance widths $y=\frac{\pi Y}{\Delta}$ (measured in units of the local
mean level spacing $\Delta(X)=1/(N\nu(X))$ of the closed system, with
$\nu(X)=\pi^{-1}(1-X^2/4)^{1/2}$ being the semicircular density of real eigenvalues) for
those resonances whose positions are within
the narrow window around the point $X$ of the spectrum is given by:
\begin{equation}\label{widres}
\langle\rho_X(y)\rangle=\frac{1}{16}
\int d\mu(\hat{Q})\mbox{Str}\left(\hat{\sigma}_\tau^{(F)}\hat{Q}\right)
\mbox{Str}\left(\hat{\sigma}_\tau\hat{Q}\right)\\ \nonumber
\prod_{a=1}^M\mbox{ Sdet}^{-1/4}\left[1-\frac{i}{2g_a}
\left\{\hat{Q},\hat{\sigma}_{\tau}\right\}\right]
\exp{\frac{i}{2}y\mbox{Str}\left(\hat{\sigma}_\tau\hat{Q}\right)}
\end{equation}
We introduced here the parameters $g_a\in [1,\infty)$ characterizing
the coupling of a system to open channels
which are expressed in terms of the eigenvalues $\gamma_a$ as
$g_a=\frac{1}{2\pi\nu(X)}
(\gamma_{a}+\gamma_{a}^{-1})$.
The integration above goes over the set of $8\times 8$
supermatrices $\hat{Q}$ satisfying the constraint $\hat{Q}^2=-1$
and some symmetry requirements, which can be conveniently resolved if one
parametrizes the manifold as: $\hat{Q}=-iZ\Lambda Z^{-1}$, with $\Lambda=diag({\bf
1,-1})$,
each ${\bf 1}$ representing $4\times 4$ unity matrix.
The symbols $Str,Sdet$ stand for the graded
trace and the graded determinant, correspondingly,
and $\left\{\hat{Q},\hat{\sigma}_{\tau}\right\}=
\hat{Q}\hat{\sigma}_\tau+\hat{\sigma}_\tau\hat{Q}$ stands
for the anticommutator.
Other $8\times 8$ supermatrices entering this expressions
 are as follows:
$$
\hat{\sigma}_{\tau}^{(F)}=
\left(\begin{array}{cc}\hat{0}_4&\hat{\tau}_3^{(F)}\\
\hat{\tau}_3^{(F)}&\hat{0}_4\end{array}\right);
\quad \hat{\sigma}_{\tau}=\left(\begin{array}{cc}0&\hat{\tau}_3\\
\hat{\tau}_3&0\end{array}\right)
$$
and $\hat{\tau}_3;\,,\hat{\tau}_3^{(F)}$ are $4\times 4$
diagonal supermatrices: $\hat{\tau}_3=\mbox{diag}\{\hat{\tau},\hat{\tau}\};
\,\,\hat{\tau}_3^{(F)}=\mbox{diag}\{\hat{0}_2,\hat{\tau}\}$.
with $ \hat{\tau}=\mbox{diag}(1,-1)$.

For the case of the matrix $\hat{\Gamma}$ being real symmetric with entries
independently fluctuating around zero and normalized in such a way that
$\lim_{N\to \infty}\mbox{Tr}  \hat{\Gamma}^2_2<\infty$ (the condition
of being "almost symmetric") the corresponding expression for the mean
density of eigenvalues is obtained from Eq.(\ref{widres}) by replacing the
product of superdeterminants in the integrand with the exponent:
$$
\exp\left\{-\frac{ v^2}{4}\mbox{Str}\left(\hat{\sigma}_\tau\hat{Q}\right)^2
\right\}
$$
where $v^2=(\pi \nu)^2\lim_{N\to \infty}\mbox{Tr}  \hat{\Gamma}^2_2$,
see the derivation in \cite{FKS3}.

The starting formulas are actually valid both for real symmetric and complex
Hermitian matrices $H$.
To extract the explicit form
of the distribution function one still has to perform the integration
over the manifold of the supermatrices $Q$. Still, it is rather
difficult calculation due to a cumbersome structure of that
manifold, and the success is mainly determined by finding an appropriate
parametrization.  For the simplest case
of matrices $\hat{H}$ being complex Hermitian the supermatrices $Q$
become block-diagonal  due to additional constraints and
 the calculation, although cumbersome, can be still done successfully
\cite{FS,FSR,FKS1} in the "standard" Efetov parametrization\cite{Efbook}.
For the case of real symmetric matrices with resulting full $8\times 8$
structure of $Q$ this is no longer the case.
Fortunately, we are able to parametrize
the matrices $Z$ as $Z_1Z_2$ in such a way, that $Z_1$ commutes
 with $\hat{\sigma}_\tau$, thus dropping out from the integrand.
We would like to stress that although we enjoyed some insights
from the recent paper on real asymmetric matrices\cite{Efnonh}, our actual
parametrization is quite different from that used in \cite{Efnonh}.
A derivation of the explicit form
of such a parametrization and especially the integration measure
turn out to be very lengthy and will be presented in details elsewhere.
Here we would like to present only the result of integrating out
the Grassmannian degrees of freedom.
\begin{equation}\label{main1}\begin{array}{c}
\displaystyle{\langle\rho_X(y)\rangle=\frac{-1}{16\pi}
\frac{\partial}{\partial y}\int_{-1}^1d\lambda f_s^{-2}(\lambda)
\int_{-\infty}^{\infty}d\lambda_1f_s(i\lambda_1)
\int_{\lambda_1}^{\infty}d\lambda_2 f_s(i\lambda_2)
\frac{(\lambda_2-\lambda_1)(2\lambda-i\lambda_1-i\lambda_2)}
{(\lambda-i\lambda_1)^2(\lambda-i\lambda_2)^2}}
\end{array}
\end{equation}
where we use the convention $s=r$ for the case of resonance statistics and
$s=R$ for that case when the antihermitian part $\hat{\Gamma}$ is a random
matrix.  Correspondingly, we defined
$$
\displaystyle{f_s(\lambda)=\frac{\exp{\lambda y}}{(1-\lambda^2)^{1/2}}
\times \left\{\begin{array}{ll}
\frac{1}{\prod_{a=1}^{M}(g_a-\lambda)^{1/2}}& \mbox{for resonances, $s=r$}\\
\exp{(v^2\lambda^2)}&
\mbox{for a random matrix, $s=R$ }
\end{array}\right.}
$$

The following important comment is appropriate here.
The integrals over $\lambda_1,\lambda_2$  as they stand
in Eq.(\ref{main1}) are not well-defined when taken literally along the real
axis due to a singularity at the points $\lambda_1=\lambda=0;\,\lambda_2=\lambda=0$.
This difficulty can be traced back to a singularity of the transformation
arising from  diagonalization of some supermatrices when coming
to a new parametrization mentioned above. A more accurate consideration shows
that the integration over $\lambda_1,\lambda_2$ should be understood in the
sense of principal value. To be precise, the correct expression is equal to
the half-sum of two integrals with integration contours encircling the
singular point $\lambda_1=-i\lambda;\, \lambda_2=-i\lambda$ from above and from below.

The expression Eq.(\ref{main1}) is exact, but still is not very convenient for
 evaluation.
Fortunately, for the physically motivated case of resonances one can
deform the contour in such a way, that it goes along the cut
$\lambda_{1,2}=-i p,\,\, p\in [1,\infty)$ for negative values of
$y$ and along a similar cut $\lambda_{1,2}=i p,\,\, p\in [1,\infty)$
for $y>0$. At the same time, all the singularities $\lambda_1=-ig_a,\, a=1,...,M$
belong to the cut in the lower half-plane. This analytic structure of the
integrand ensures vanishing of the density for $y>0$ as expected, whereas
for $y<0$ one obtains:
\begin{equation}\label{main2}\begin{array}{c}
\displaystyle{\rho_X(y<0)=\frac{1}{4\pi}
\frac{\partial^2}{\partial y^2}\int_{-1}^1d\lambda (1-\lambda^2)
\int_{1}^{\infty} \frac{dp_1}{(p_1^2-1)^{1/2}}
\int_{1}^{p_1} \frac{dp_2}{(p_2^2-1)^{1/2}}
\frac{(p_1-p_2)e^{+y(p_1+p_2-2\lambda)}}
{(\lambda-p_1)^2(\lambda-p_2)^2}}\\
\times \displaystyle{\chi_1(p_2)\chi_2(p_1)\prod_{a=1}^M
\frac{(g_a-\lambda)}{(|g_a-p_1||g_a-p_2|)^{1/2}}}
\end{array}
\end{equation}
where the functions $\chi_{1,2}(p)$ are defined as follows:
$$
\displaystyle{ \chi_1(p)=\left\{\begin{array}{l}
1\quad \mbox{for}\,\, g_{4k}\le p\le g_{4k+1} \\
-1\quad \mbox{for}\, g_{4k+2}\le p\le  g_{4k+3} \\ 0\quad \mbox{otherwise}
\end{array}\right. \quad \quad
\chi_2(p)=\left\{\begin{array}{l}
1\quad \mbox{for}\,\, g_{4k+1}\le p\le g_{4k+2} \\
-1\quad \mbox{for}\, g_{4k+3}\le p\le  g_{4k+4} \\ 0\quad \mbox{otherwise}
\end{array}\right.}
$$
where $k=0,1,...,[M/4]$ and we used a convention: $1\equiv g_0\le g_1\le
...\le g_M<g_{M+1}=\infty$.

The formulas Eqs.(\ref{main2}) provide us with the most general
explicit analytical expression for the density of complex
resonances for a chaotic quantum system with preserved time-reversal
invariance. As such
they constitute the main result of the present paper.

We see, that in general these formulas are still quite cumbersome.
One can infer from them various known limiting cases
considered earlier. For example, we can consider the case of weak coupling
to continuum when all $g_a\gg 1$. By noticing that the integration
over $p_2$ in Eq.(\ref{main2}) is dominated in this case by $p_2\sim 1$, whereas
 $p_1\ge g_1>>1,\lambda,p_2$ and taking into account also that
typical values of $y$ are of the order of $g_1^{-1}<<1$ one can evaluate
the integrals over $p_2,\lambda$ to the leading order.
The result is:
\begin{equation}\label{AL}
\displaystyle{\rho_X(y<0)=\frac{1}{\pi}\prod_{a}g_a^{1/2}\int_{1}^{\infty}
\frac{dp_1}{\prod_{a}|g_a-p_1|^{1/2}} e^{y p_1}\chi_2(p_1)}
\end{equation}
which coincides with the distribution derived
recently by Alhassid and Lewenkopf\cite{AL} by a perturbative treatment,
up to alternating signs in the factor $\chi_2(p)$ which is an apparent
misprint in their paper.

Relying on earlier experience with resonance statistics in systems with
broken time-reversal invariance\cite{FS,FSR}, one might expect that the
distribution  simplifies drastically for the case of statistically equivalent
channels:
$g\equiv g_1=g_2=...=g_M$. Surprisingly, this seems to be not the case for
the present model.
Actually, we find it difficult to perform such a limit
explicitly for arbitrary number of open channels, and even for
$M=4$ the result turns out to be quite cumbersome. Below we present the
formula for  one and two open channels $M=1,2$:
\begin{equation}\label{M12}
\rho_X(y<0)=\frac{1}{4\pi}
\frac{\partial^2}{\partial y^2}\int_{-1}^1d\lambda (1-\lambda^2)
\displaystyle{e^{-2\lambda y} F_M(\lambda,y)}
\end{equation}
where for $M=1$ we found
$$
F_1=(g-\lambda)\int_g^{\infty}
dp_1 \frac{e^{y p_1}}{(p_1^2-1)^{1/2}(\lambda-p_1)^2
(p_1-g)^{1/2}}
\int_{1}^{g}dp_2 \frac{e^{y p_2}}{(p_2^2-1)^{1/2}}
\frac{(p_1-p_2)}
{(\lambda-p_2)^2(g-p_2)^{1/2}}
$$
and the case $M=2$ is the simplest one:
$$
F_2=\frac{\pi}{(g^2-1)^{1/2}}
\int_{1}^{g}dp_2 \frac{1}{(p_2^2-1)^{1/2}}
\frac{e^{y(p_2+g)}}{(\lambda-p_2)^2}
$$

The figures Fig.1 and  and Fig.2 show that
the resulting expression favourably agree with the results of direct numericaldiagonalization of the corresponding complex matrices.

Despite the fact, that the general resonance widths distribution for the
case of preserved TRS turns out to be much more complicated
  than for the case of broken TRS, both expressions share many
important features. For example, evaluating the first moment of the
distribution Eq.(\ref{main2}) for $M=2$ exactly, one arrives at the following
expression for the mean resonance widths:

\begin{equation}\label{Simonius}
\frac{\langle\Gamma\rangle}{\Delta}=-
\frac{1}{\pi}\ln{\frac{g-1}{g+1}}
\end{equation}
This formula is well known in nuclear physics as Moldauer-Simonius
relation and follows from the unitarity of the scattering matrix\cite{Mol}.
Actually, a simlar relation must be satisfied for an arbitrary number $M$ of open
channels. It was indeed found to be the case for
systems with broken TRS\cite{FSR}. Unfortunately, it is quite difficult to
verify the Moldauer-Simonius relation directly from the general distribution
Eq.(\ref{main2}).

The logarithmic divergency at $g_a=1$ signals on the powerlaw tail
behaviour $\rho_X(|y|\gg 1)\propto y^{-2}$ typical for this (a so-called
"perfect coupling") case. It is instructive to perform the limit $g\to 1$
directly in  Eq.(\ref{M12}) for $M=2$ and find:
\begin{equation}\label{M2g1}
\rho_X(y<0)=\frac{1}{4y^3}\left(2y+1+(2y-1)e^{4y}\right)
\end{equation}
where such a behaviour is self-evident.

Actually, it is interesting to compare this expression
with the $M=1$ case for systems with broken TRS\cite{FS}:
\begin{equation}\label{M2g11}
\rho_X(y<0)=\frac{1}{2 y^2}\left(1+(2y-1)e^{2y}\right)
\end{equation}
The reason is that in the regime of isolated resonances $g\gg 1$
the M-channel system with broken TRS possesses exactly the same
widths distribution as the $2M$ channel system with unbroken TRS.
A similar correspondence also holds in the limit of many open channels
$M\gg 1$. Moreover, before the exact formula Eq.(\ref{M2g1}) was available,
an attempt was made\cite{Seba} to compare $M=1$ formula
Eq.(\ref{M2g11}) valid for systems with broken TRS with numerically found
resonance widths in $M=2$ chaotic TRS-preserving microwave cavity.
The correspondence was reported to be quite satisfactory.

We see, that both formulas indeed give exactly the same behaviour
for the resonances whose widths exceed the typical separation between them:
$\rho_X(|y|\gg 1)\approx 1/2y^2$. However, in the region of very narrow
resonances the behaviour is slightly different: $\rho_X(y\to 0)=1$ for
broken and $\rho(y\to 0)=4/3$ for unbroken TRS. This fact shows the
limitations of the mentioned correspondence.

As to the random-perturbation case ($s=R$ in eq.(\ref{main1}), the presence
of the Gaussian factors in
the integrand of Eq.(\ref{main1}) prevents one from deforming the contour to
the complex plane. At the moment, we are unable to simplify the expression
further. However, one can still infer
 that in the perturbative case $v\ll 1$ the function $\rho_X(y)$
reduces to a simple Gaussian, whereas in the opposite limiting case
$v\gg 1$ one indeed arrives at the "elliptic law" well-known to hold
for strongly non-Hermitian random matrices\cite{Gir,neural}.

The authors are much obliged to
B.A.Khoruzhenko for useful communications.
The work was supported by SFB 237 "Unordnung und grosse Fluktuationen"
and in part by RFBR Grant No. 96-02-18037-a.




\begin{figure}
\epsfxsize=0.9\hsize
\epsffile{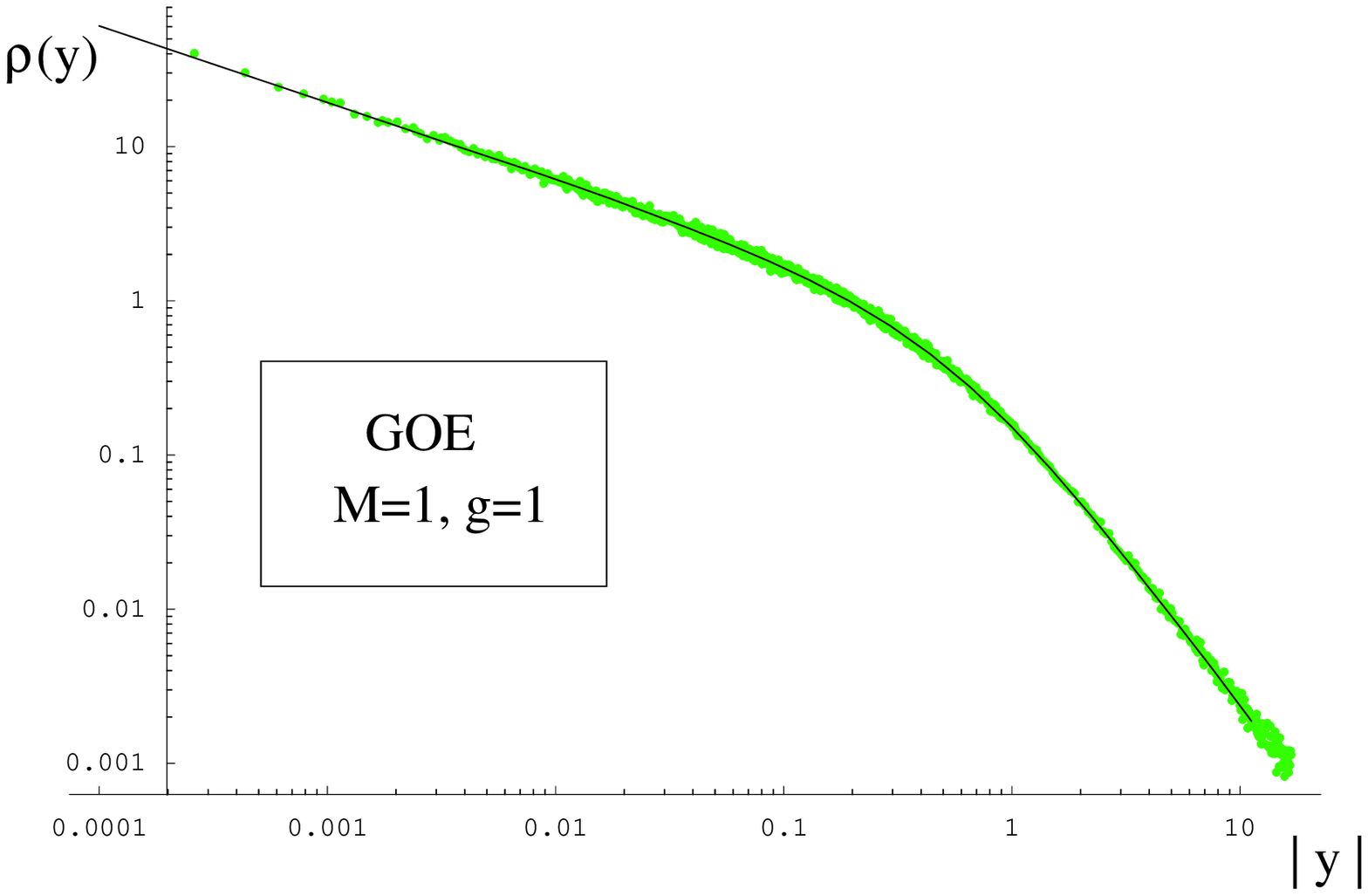}

\vspace{0.1\hsize}
\caption{
\label{goem1g1}
Comparison between the predictions of our formula Eq.(\ref{M12}) for one channel case M=1 at "perfect coupling" g=1 and the results of direct numerical diagonalization of $K=10000$ complex matrices of the size $400\times 400$.}
\end{figure}

\newpage
\begin{figure}
\epsfxsize=0.9\hsize
\epsffile{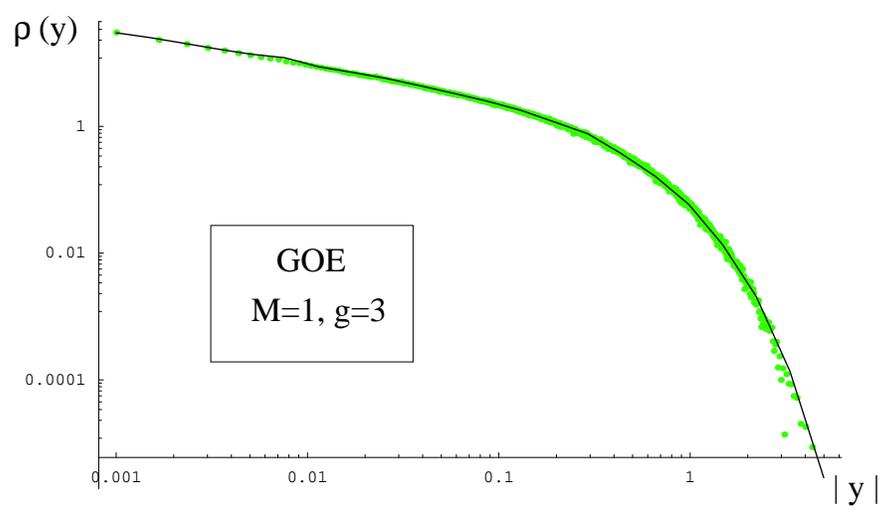}

\vspace{0.1\hsize}
\caption{
\label{goem1g3}
Same as Fig.1 for M=1, g=3.
}
\end{figure}

\end{document}